\documentstyle[aps,prl,preprint]{revtex}
\begin{document}
\title { Fermi surface evolution in underdoped cuprates}
\author{Andrey V. Chubukov$^{1,2}$, Dirk K. Morr$^{1}$ and Konstantin A.
Shakhnovich$^{3}$}
\address{
$^{1}$Department of Physics, University of Wisconsin, Madison, WI 53706\\
$^{2}$P.L. Kapitza Institute for
Physical Problems, Moscow, Russia\\
$^{3}$ University of Chicago, Chicago, IL}
\date{today}
\maketitle

\begin{abstract}
We consider a  two-dimensional
 Fermi liquid coupled to low-energy commensurate spin fluctuations.
 At small coupling, the hole Fermi surface is large and
centered around $(\pi,\pi)$.
 We show that as the coupling increases, the shape of the
Fermi surface undergoes a substantial evolution,
 and at strong coupling, the hole Fermi surface
consists of small pockets centered at $(\pm \pi/2, \pm \pi/2)$.
At intermediate couplings, there exists a large hole Fermi surface
centered at $(\pi,\pi)$ as well as four hole pockets, but the
quasiparticle residue is small everywhere except for the pieces
 of the pockets which face the $\Gamma$ point.
 The relevance of these results to recent
photoemission experiments in $YBCO$ and $Bi2212$ systems,
 and their relation to the Luttinger theorem are
discussed.
\end{abstract}


A.C: It is a great pleasure to present a paper for a Festschrift in
honor of David
Pines. David's contributions to many fields in theoretical physics are
well recognized throughout the world, as well as his charming personality.
David, we wish you many  more successful years in Many-Body Theory.
******************************************************************************
The problem of fermions interacting with low-energy magnetic fluctuations
has attracted a considerable interest over the past few years particularly
in connection with high-$T_c$
superconductivity (Hertz 1976, Ioffe and Larkin
1989,  Kampf and Schrieffer 1990,
 Ioffe and Kotliar 1990, Lee and Nagaosa 1992,
Millis 1993, Monthoux
and Pines 1994,  Sachdev, Chubukov and Sokol 1995,
Altshuler, Ioffe and  Millis 1995,
Sachdev and Georges 1995, Wen and Lee  1995).  The vast majority of
experimental results for cuprates indicate
 that the superconducting gap is highly
anisotropic and its dominant part possesses
$d_{x^2 - y^2}$ symmetry. The process which yields  this kind of
pairing is the exchange of nearly antiferromagnetic spin fluctuations.
Photoemisson experiments performed at optimal doping (Liu, Veal, Paulikas,
Downey and Shi  1992, Liu, Veal, Paulikas, Downey, Kostic, Fleshler,
Welp, Olson, Wu, Arko and Joyce  1992,
 Dessau, Shen, King, Marshall, Lombardo, Dickinson, Loeser, DiCarlo, Park,
Kapitulnik and Spicer 1993,
Ma, Quitmann, Kelley, Almeras, Berger, Margaritondo and Onnelion 1995)
show that electrons
possess a large, Luttinger-type Fermi surface whose area scales as $1 + x$
where $x$ is the doping concentration.
For this Fermi surface, D. Pines and Ph. Monthoux (Monthoux et al. 1994)
have performed a number of
strong coupling calculations of $T_c$ within the Eliashberg formalism, and
found $T_c$ consistent with experiments.

There are however several issues related to
 the magnetic mechanism for superconductivity,
which are still subjects of debates. Naively, one would expect that the
magnetic mechanism should be the  stronger, the closer the system is
to the antiferromagnetic instability.  Experimentally, however, $T_c$ has a
maximum at  much larger doping concentrations than those where long-range
magnetic order has been observed. Moreover, for a system with long-range order,
one can study the electronic excitations and effective pairing interaction
within the SDW technique (Schrieffer, Wen and Zhang 1989, Chubukov and
Frenkel 1992). At large Hubbard $U$, the
ordered state possesses two bands of quasiparticles with $Z \approx 1/2$
separated by a gap, $2\Delta \approx U$. Near half-filling, the
valence band is nearly fully occupied, and the conduction
 band is empty. Accordingly, the hole Fermi surface has a {\it small} area
which scales as $x$ rather than $1+x$.  The photoemission studies of
$Sr_2CuO_2Cl_2$ (Wells, Shen, Matsuura, King, Kastner, Greven and Birgeneau
1995)
have demonstrated that the valence band maxima are at
$(\pi/2,\pi/2)$ and symmetry related points. This is consistent with the
theoretical studies which yield hole pockets
at $(\pm \pi/2, \pm \pi/2)$ near half-filling.
For this kind of Fermi
surface, there are simply no low-energy quasiparticles in the directions of
 momentum space, where the $d_{x^2 - y^2}$ gap has the largest amplitude.
Moreover, the interaction vertex between fermions and spin fluctuations obeys
the Adler principle, i.e., it vanishes when the magnon momentum
equals $(\pi,\pi)$. As a result, the full scattering potential between
fermions,
mediated by the exchange of spin fluctuations, does not have a peak at
$(\pi,\pi)$ and therefore one cannot get $d_{x^2 - y^2}$
pairing from the magnetic mechanism (Schrieffer 1994).

In a mean-field SDW approach, the gap, $2 \Delta$,
between conduction and valence bands is proportional to the sublattice
magnetization, i.e., it
 vanishes when long-range order
disappears. However, this
is actually an artifact of the mean-field treatment.
The easiest way to see this is to switch on a small, but finite temperature
and compute lowest-order
temperature corrections to the mean-field formulae. The corrections to
the sublattice magnetization
are logarithmically divergent which
is an obvious consequence of the Mermin-Wagner theorem.
At the same time, the divergent terms in the corrections to $\Delta$
are cancelled out.
 As a result, at finite $T$, $2\Delta$ remains approximately equal to $U$,
 while the sublattice magnetization
vanishes. This result is expected in some sense because
the existence of two narrow bands with quasiparticle residue $Z \approx
1/2$, separated by an energy gap $U$, is similar to what one obtains by
just considering the atomic limit of the Hubbard model near half-filling.
{}From this perspective, the role of strong antiferromagnetic fluctuations
 is to transform the two Hubbard levels into {\it
coherent} bands of quasiparticles~\footnote[1]{Without strong magnetic
fluctuations,  the two bands which emerge from the Hubbard
levels are likely to be mostly incoherent, and there also appears, at finite
$t$, a coherent band at about $U/2$. This result was found in infinite-$D$
studies of the Hubbard model (Rozenberg, Kotliar, Kalueter, Thomas, Rapkine,
 Honig and Metcalf 1995).}.

The consideration above indicates that one can obtain an ``almost'' SDW
description of the electronic states, with $2\Delta \approx U$, even without
long-range order.
This is consistent
with QMC studies (Bulut, Scalapino and White 1994,
Preuss, Hanke and von der Linden 1995)
 which show that the
density of states in a paramagnetic phase
near half-filling has two distinct peaks roughly separated by $U$.
However, as the system moves away from half-filling,
the spectral weight transferes
from the  upper band into the lower band, and near optimal
doping there exists just one coherent band of quasiparticles.
This  implies that the
features of an SDW solution appear gradually as the system moves from
optimal doping to half-filling.

In the present paper, we analyse how the system acquires the properties of the
SDW solution. We consider
 a simple model in which electrons are coupled to strong
antiferromagnetic fluctuations by
\begin{equation}
{\cal H}_{int} = g \sum_{k,q,\alpha,\beta} c^{\dagger}_{k+q, \alpha} {\vec
\sigma}_{\alpha,\beta} c_{k,\beta} {\vec S}_{q}
\label{intham}
\end{equation}
where $g$ is the coupling constant, and $\sigma_i$ are the Pauli matrices.
We assume that the weak-coupling limit describes optimal doping
and model the system's evolution towards
half-filling by increasing the strength of the coupling constant $g$.
The partial
justification for this assumption follows from the diagrammatic
consideration of
the near atomic limit of the Hubbard model which will be presented elsewhere.
 We will also assume
 that the original $SU(2)$
symmetry is extended to $SU(N)$, substitute Pauli matrices by $N^2 -1$
traceless $SU(N)$ generators,  and perform the calculations
for $N \rightarrow \infty$. In this limit, the mean-field SDW description
becomes exact.
Finally, we will  assume that the dynamical spin susceptibility,
$\chi (q,\omega)$
is peaked at $Q = (\pi,\pi)$ and its low-energy part has the
form $\chi^{-1}_{i i} (q, i\omega_m) =
A^{-1} (\delta^2 + v^2_s (q - Q)^2 + \omega^2_m + 2 \gamma
|\omega_m|)$.
We will consider $\delta$ as a parameter which does not vary with the coupling
strength.
As a further simplification, we will assume that this form of the
susceptibility holds when $v_s |q-Q|$ and $|\omega|$ are both smaller than
some cutoff scale, $C \ll W$, where $W$ is the fermionic bandwidth.
 At higher
energy scales, we simply set $\chi_{i i} (q,\omega) =0$. This sharp cutoff
however is only used here to simplify the calculations - we will see that
in order to obtain a strong Fermi surface evolution with increasing $g$ we only
have to assume that the spin susceptibility gets substantially reduced at the
 scales which are larger than $W$.  The overall factor $A$ is determined from
$\int d{\vec q} d \omega~\chi_{i i} (q, \omega) = {\hat S}^2/(N^2-1)$.

We now proceed to the calculations of the quasiparticle self-energy.
We will first present the results for $\gamma =0$ when a
full analytical calculation is possible, and then show how the results are
modified due to finite damping.

 The second-order self-energy contains the Green function of an
intermediate state with momentum $k+q$.
 Since $C \ll W$,
 we can expand the fermionic energy
$\epsilon_{k+q}$ as
$\epsilon_{k + q} \approx \epsilon_{k +Q} + {\vec
v}_{k+Q} {\vec {\bar q}}$, where ${\bar q} = q -Q$.
 Within this approximation, we obtained the analytical
expression for $\Sigma (k,\Omega)$
for arbitrary ratio of $ a_k =|v_{k+Q}|/v_s$.
It turns out, however,
that the mechanism of the Fermi surface
evolution is virtually insensitive to the value and
momentum dependence of $a_k$.
The results for arbitrary $a_k$ will be presented in a separate
publication, here we just set
 $a_k =1$ in which case the expression for the self-energy has the
simple form~\footnote[2]{In performing
the integration, we actually set ${\vec v}_{k+Q}
= v^x_{k+Q}$, integrated over ${\bar q}_y$
from $-\infty$ to $+ \infty$ and over the
circle $v^2_{k+Q} {\bar q}_x^2 + \omega_m^2 < C^2$.
 However, we also checked that the
results change very little if we also restrict the integration over
${\bar q}_y$ to $|v_{k+Q} {\bar q}_y| < C$.}
\begin{equation}
\Sigma (k, \Omega_m) = - g_e^2~\frac{1}{{\bar \epsilon}_{k+Q} - i
\Omega_m}
\label{self-ena}
\end{equation}
for  ${\bar \epsilon}_{k+Q}^2 + \Omega_m^2 \geq C^2$, and
\begin{equation}
\Sigma (k, \Omega_m) = - g_e^2~\frac{1}{{\bar \epsilon}_{k+Q} - i
\Omega_m}~\frac{\sqrt{{\bar \epsilon}_{k+Q}^2 + \Omega_m^2 + \delta^2} -
\delta}{\sqrt{C^2 + \delta^2} -\delta}
\label{self-enb}
\end{equation}
for  ${\bar \epsilon}_{k+Q}^2 + \Omega_m^2 < C^2$. Here we defined
${\bar \epsilon}_k = \epsilon_k - \mu$, and $g_e = (2/N)^{1/2} g |{\hat S}|$.
The location of the Fermi surface follows from $G^{-1} (k,0) = -
({\bar \epsilon}_k + \Sigma (k,0)) =0$.

The relevant issue for our consideration is how the chemical
potential $\mu$ changes with $g_e$. The value of $\mu$ is obtained from the
condition that $2 \int d {\vec k} d \omega ~G (k,\omega) = N/V$. We solved this
equation numerically and present the result in Fig. 1. We found
 that $\mu$ remains nearly the same as for free
particles roughly
up to $g_{e} = g^{(3)}_3$, when pockets are formed (see below),
 and then starts growing and approaches $g_e$ at large couplings.

It is instructive to consider the large $g_e$ limit in more detail. As $|\mu|$
is large in this limit, the condition
$\Omega^2_m + {\bar \epsilon}^2_{k+Q} > C^2$ is satisfied for all momenta and
Matsubara frequencies (though not for all real frequencies!),
 and the self-energy is given by
(\ref{self-ena}). For the full Green function we then immediately obtain
\begin{equation}
G (k, \Omega_m) = \frac{i\Omega_m - {\bar \epsilon}_{k+Q}}{(i \Omega_m - E_1
(k)) (i \Omega_m - E_2 (k))},
\label{sdw-like}
\end{equation}
where $E_{1,2} (k)$ are given by
\begin{equation}
E_{1,2} (k) = \frac{{\bar \epsilon}_{k+Q} + {\bar \epsilon}_{k}}{2} \pm
\left(g_e^2 + \left(\frac{{\bar \epsilon}_{k+Q} - {\bar
\epsilon}_{k}}{2}\right)\right)^{1/2}.
\label{E12}
\end{equation}
Performing now an analytical continuation to real frequencies, we find that
 the full Green function  has two poles at $ \Omega = E_{1,2} (k) \approx
|\mu| \pm g_e$.
It is easy to check that for both solutions the condition for using
(\ref{self-ena}) for the self-energy (which, for real frequencies,
 is, $|\Omega^2 - {\bar \epsilon}^2_{k+Q}| >C^2$) is satisfied.
These two solutions obey $E_{1,2} (k) = E_{1,2} (k+Q)$ and
have exactly the same form as the mean-field solutions of the ordered
SDW state with the effective coupling $g_e$ playing the role of the SDW gap,
$\Delta$.
 This exact equivalence with the SDW formulae
is indeed the result of restricting with only the leading terms in the
expansion in $C/g_e$
 - in the absence of a broken symmetry, $k$ and $k+Q$ are {\it not}
equivalent points in the Brillouin zone, and in this sense, the quasiparticle
bands at $E_{1,2}$ are only the
precursors of the conduction and valence
bands of the ordered state. Further, it follows from (\ref{sdw-like}) that
the quasiparticle
residue for each of these two solutions is $Z = 1/2 +O(t/g_e)$.
Clearly then, the hole Fermi surface is small and encloses the
area which scales as $x$ rather
than $1+x$ as was the case for small $g_e$.

We now present our results for the Fermi surface evolution with
varying $g_{e}$ (Fig. 2).
We found that the evolution begins with the ``hot spots'',
 where ${\bar \epsilon}_k = {\bar \epsilon}_{k+Q} =0$. It is clear
 from (\ref{self-enb}) that as $\Sigma ({\bar \epsilon}_{k+Q} =0, \Omega =0)
=0$, the location of the ``hot spots'' is not
shifted by self-energy corrections. However, at $g_e = g^{(1)}_e = (2 \delta
(\sqrt{C^2 + \delta^2} - \delta))^{1/2}$, the Fermi velocity at these points
turns to zero. At $g_e > g^{(1)}_e$, each of the ``hot spots'' is split into
three: one  is still at ${\bar \epsilon}_k = 0$, while
the other two are located at ${\bar \epsilon}_k = \pm (g^2_e /(\sqrt{C^2 +
\delta^2} - \delta))(1 - (g^{(1)}_e/g_e)^2)^{1/2}$ at $g_e <C$, and
at ${\bar \epsilon}_k = \pm g_e$ when $g_e > g^{(2)}_e =C$ (Fig. 2b).
Besides, at $g_e >g^{(2)}_e$, there appear (for $a_k =1$) nested
 pieces of the Fermi surface located
between points $D$ and $D^{\prime}$ in Fig. 2c. For $g_e$ only
slightly larger than $g^{(2)}_e$, both $D$ and $D^{\prime}$ are located near
the magnetic Brillouin zone boundary. As $g_e$ increases further, these points
move apart and approach the Brillouin zone diagonal. Finally, at $g_e =
g^{(3)}_e = C (1 +  8t(|\mu| -C)/C/(2t + \sqrt{4t^2 - 4|t^{\prime}|(|\mu|
-C)})$, the $D$ and $D^{\prime}$ points from neighboring ``hot spots''
merge and the system undergoes a topological transition when the
singly-connected hole Fermi surface splits into hole pockets centered at
$(\pm \pi/2, \pm \pi/2)$ and the large hole Fermi surface (Fig. 2d).
 As $g_e$ increases further, the large Fermi surface
shrinks (Fig. 2e) and eventually disappears, and at even larger $g_e$, the
Fermi surface consists of just four hole pockets (Fig. 2f).

We further discuss how the quasiparticle residue along the Fermi surface
changes with the coupling strength.  At small $g_e$, the residue is nearly
$k-$independent and is close to $Z=1$. At large $g_e$, $Z$ along the hole
pockets
is again $k-$independent and is approximately equal to $1/2$. At intermediate
$g_e$, however, we found a strong momentum dependence of $Z$.
 In Fig. 2e we present our
results for $Z$ when large and small Fermi surfaces coexist. We see that
the quasiparticle residue along the large Fermi surface is very small which
makes
its experimental observation problematic. The residue along the hole pocket is
also very anisotropic and is relatively
close to $1$ only in a momentum region which faces the $\Gamma$ point. This is
consistent with the experimental observations by R. Liu et al.~who reported
a small Fermi surface in $YBa_2Cu_3O_{6.3}$, but could only detect
the Fermi surface crossing between the $\Gamma$ and $(\pi/2,\pi/2)$ points (Liu
et al. 1992).

We now discuss how the results above change when we include
the damping of spin fluctuations. We found that the general
scenario of the Fermi surface evolution
does not change, but the value of $g^{(1)}_e$ where each
 ``hot spots'' splits into three is in fact
substantially larger than in the absence of damping.
Specifically, we found (still, assuming for simplicity that
$a_k =1$) that $g^{(1)}_e (\gamma) =
g^{(1)}_e (\gamma =0) \Psi (\gamma/\delta)$ where
$\Psi (x) = 1 + x/(3\pi) +O(x^2)$ for $x \ll 1$, and $\Psi (x) = (16 \log
x/(\pi x))^{-1/2}$ for $x \ll 1$.  Near optimal doping (which corresponds to
weak/intermediate couplings in our consideration), spin fluctuations are
overdamped, i.e., $\gamma \gg \delta$. In this case, $g^{(1)}_e \propto
(\gamma C/\log \gamma/\delta)^{1/2}$
 which is substantially larger than $g^{(1)}_e
\propto (\delta C)^{1/2}$ which we obtained in the absence of damping.
 Though the actual numbers depend on
the details of the band structure and specific predictions about $v_s$, etc, it
is essential that this critical value {\it does not} contain $\delta^{1/2}$
 as an overall factor.

Now about vertex corrections.
We performed the same lowest-order computations
as for the self-energy. At small and moderate $g_e$, the result strongly
depends on the momenta of the external particles. In particular, it has been
shown (Altshuler et al.  1995, Chubukov  1995) that
if we set the
momentum in the spin propagator $q=Q$, and choose the incoming and outgoing
fermions to be right
at the original ``hot spots''
  (such that both fermions are simultaneously at the Fermi surface),
then vertex corrections in fact {\it increase} the pairing
potential and favor
$d-$wave pairing.
However, if we choose fermions to be at the new ``hot spots'' which
appear at $g_e > g^{(1)}_e$, then the vertex corrections have opposite
sign
and decrease the pairing potential.
At large $g_e$, only a single new ``hot spot''
survives. In this limit, we found (for $a_k =1$ and to leading order in
$C/g_e$) that the full vertex function is
\begin{equation}
g^{full}_e (k,\Omega_m) = g_e~\frac{(i\Omega_m - E_1 (k))
(i\Omega_m - E_2 (k))}{(i\Omega_m - {\bar \epsilon}_{k})(i\Omega_m
- {\bar \epsilon}_{k+Q})}
\label{vertex}
\end{equation}
where $E_{1,2}$ are given by (\ref{E12}). This form of the vertex
function is the same as in the mean-field SDW solution for the
ordered state. An analytical continuation of (\ref{vertex}) to real frequencies
requires care as (\ref{vertex}) is valid only when
$|\Omega^2 - {\bar \epsilon}^2_{k}| > C^2$ and
$|\Omega^2 - {\bar \epsilon}^2_{k+Q}| > C^2$.
At $\Omega = E_{1,2} (k)$, these two conditions are satisfied, and we
see that the full vertex function
vanishes at resonance.
Indeed, in the absence of the symmetry breaking,
vanishing of $g^{full}_e$  is again
a result of neglecting the
subleading terms. Nevertheless, we see that at $g_e \gg C$,
 vertex corrections substantially reduce the effective
pairing potential between fermions mediated by the exchange of spin
fluctuations.

Now consider higher-order diagrams. Here the key point is that at large $g_e$,
our results are the same as in the
SDW description. Meanwhile, we found that
the full mean-field SDW solution is
given by the second-order self-energy diagram. Higher-order terms do not
contribute because of a precise cancellation (at $n \rightarrow \infty$)
between self-energy and vertex corrections to the second-order self-energy
term. In our case SDW solutions
 are reproduced only to lowest-order in $C/g_e$, i.e. there is no
 complete cancellation. However, the higher-order
self-energy terms are still small, at $g_e \gg C$,
compared to the  second-order self-energy and
can be neglected.
 At intermediate $g_e \sim C$, the restriction with
only the second-order diagram is, strictly speaking, unjustified, but we expect
that the scenario for the Fermi surface evolution will
remain at least qualitatively the same as the one described above.

Finally, some considerations on the Luttinger theorem.
It is clearly violated at
large $g_e$.
The source of this violation will be discussed in a separate publication,
here we only notice that the key point of the Luttinger theorem is the proof,
order by order in perturbation theory,
that $I = \int d{\vec k} d \Omega ~G(k,\Omega)~ d \Sigma (k,\Omega)/d\Omega
=0$.
Meanwhile, for the SDW state, $I$ is actually finite, though a formal
application
of Luttinger's reasoning still yields $I=0$ to all orders in perturbation
theory.
The point is that the nonzero contribution
to $I$ comes solely from the region of $(k,\Omega)$ where
the Luttinger expansion parameter
$ \Lambda =\Delta^2 |G_0 (k,\Omega) G_0 (k+Q,\Omega)|>1$, and
perturbation series do not converge.  We computed $I$ and also the area
of the Fermi surface with our self-energy. The results for the Fermi surface
area are presented in Fig. 1b.
Though at intermediate $g$,
higher-order diagrams cannot be neglected and we cannot exactly reproduce the
Luttinger theorem even when it should be valid, our
 numerical data indicate that the area
stays nearly the same as for free fermions roughly up to
$g^{(3)}_e$ when
 the hole pockets get separated
from the rest of the Fermi surface. We believe, though at the moment have no
proof, that the Luttinger theorem ceases to work above $g^{(3)}_e$.

It is our pleasure to thank S. Chakravarty, E. Dagotto,
A. Finkelstein, D. Frenkel,
L. Ioffe, R. Joynt, A. Kampf, R. Laughlin, A. Moreo, P. Lee,
A. Millis, M. Onellion, D. Pines, A. Ruckenstein, S. Sachdev, D. Scalapino,
B. Shraiman, A. Sokol and Q. Si for numerous discussions and comments.
A.C. is an A.P. Sloan fellow.

\newpage

{\center  {\Large References \\[1cm]}}

\begin{enumerate}

\item B. L. Altshuler, L. B. Ioffe, and A. J. Millis, preprint

\item N. Bulut, D. J. Scalapino, and S. R. White, Phys. Rev. Lett. {\bf 73},
748 (1994)

\item A. V. Chubukov and D.M.Frenkel, Phys. Rev. B {\bf 46}, 11884 (1992)

\item A. V. Chubukov, Phys. Rev. B {\bf 52}, R3847  (1995)

\item D.S. Dessau, Z.-X. Shen, D. M. King, D. S. Marshall, L. W. Lombardo, P.
H.
Dickinson, A. G. Loeser, J. DiCarlo, C.-H. Park, A. Kapitulnik,
and W. E. Spicer, Phys. Rev. Lett. {\bf 71}, 2781 (1993)

\item J.A. Hertz, Phys. Rev. B {\bf 14}, 1165 (1976)

\item L.B. Ioffe and A.I. Larkin, Phys. Rev. B {\bf 39}, 8988 (1989)

\item L.B. Ioffe and G. Kotliar, Phys. Rev. B {\bf 42}, 10348 (1990)

\item L.B. Ioffe, V.Kalmeyer and P.B. Wiegmann, Phys. Rev. B {\bf 43}, 1219
(1991)

\item L.B. Ioffe and A. Millis, preprint

\item A.P. Kampf and J.R. Schrieffer, Phys. Rev. B {\bf 42}, 7967 (1990)

\item C. Kane, P. Lee and N. Read, Phys. Rev. {\bf 39}, 6880 (1989)

\item P.A.Lee and N. Nagaosa, Phys. Rev. B {\bf 46}, 5621 (1992)

\item R. Liu, B. W. Veal, A. P. Paulikas, J. W. Downey, and H. Shi,
Phys. Rev. B {\bf 45}, 5614 (1992)

\item R. Liu, B. W. Veal, A. P. Paulikas, J. W. Downey, P. J. Kostic, S. Fleshl
er, U. Welp, C. G. Olson, X. Wu, A. J. Arko, and J. J. Joyce,
Phys. Rev. B {\bf 46}, 11056 (1992)

\item Jian Ma, C. Quitmann, R. J. Kelley, P. Almeras, H. Berger,
G. Margaritondo, and
M. Onnelion, Phys. Rev. B {\bf 51}, 3832 (1995); Jian Ma,
P. Almeras, R. J. Kelley, H. Berger, G. Margaritondo, X. Y. Cai,
 Y. Feng, and M. Onellion, Phys. Rev. B {\bf 51}, 9271 (1995)

\item A.J. Millis, Phys. Rev. B {\bf 48}, 7183  (1993)

\item P. Monthoux and D. Pines, Phys. Rev. B {\bf 49}, 4261 (1994)

\item R. Preuss, W. Hanke, and W. von der Linden, Phys. Rev. Lett. {\bf 75},
 1344 (1995)

\item M. Rozenberg, G. Kotliar, H. Kalueter, G. A. Thomas, D. H. Rapkine, J. M.
Honig, and P. Metcalf, Phys. Rev. Lett. {\bf 75}, 105 (1995)

\item S.  Sachdev, A.V.  Chubukov and A. Sokol, Phys. Rev. B {\bf 51}, 14874
(1995).

\item S. Sachdev and A. Georges, Phys. Rev. B {\bf 52}, 9520 (1995).

\item J.R.Schrieffer, X.G.Wen, and S.C.Zhang, Phys. Rev. B {\bf 39}, 11663
(1989).

\item J. R. Schrieffer, J. Low Temp. Phys. {\bf 99}, 397 (1995).

\item B. I. Shraiman and E.D. Siggia, Phys. Rev. Lett. {\bf 61}, 467 (1988).

\item X-G Wen and P.A. Lee, preprint.

\end{enumerate}

\newpage

\begin{figure}
\caption {{\em (a)} The chemical potential as a function of the coupling,
$g_e$. We used (in units of $t$), $\delta
=0.1,~C=0.3,~t^{\prime} =-0.45$, $x =0.1$.
For free fermions, $\mu =\mu_0 \approx -1.16$.
The arrow indicates the value of $g^{(3)}_e$
when hole pockets are formed. {\em (b)} The area of the occupied states vs
$g_e$. The dashed line is the area for free fermions.}
\end{figure}

\begin{figure}
\caption {{\em (a) -(f)} Fermi surface evolution with increasing $g_e$.
The parameters are the same as in Fig. 1. For these parameters,
$g^{(1)}_e = 0.21,~g^{(2)}_e = 0.3$, and $g^{(3)}_e \approx 0.82$.
The figures are for $g_e =0~ g_e = g^{(2)}_e > g^{(1)}_e,
{}~g^{(3)}_e> g_e >g^{(2)}_e, ~g_e \geq g^{(3)}_e,~ g_e > g^{(3)}_e$,
and $g_e \gg g^{(3)}_e$, respectively. In {\em (c)}, the
nested pieces of the Fermi surface
are shown in bold.  In Fig. {\em (e)}, we also presented the values
of the quasiparticle residue along the Fermi surface.}
\end{figure}

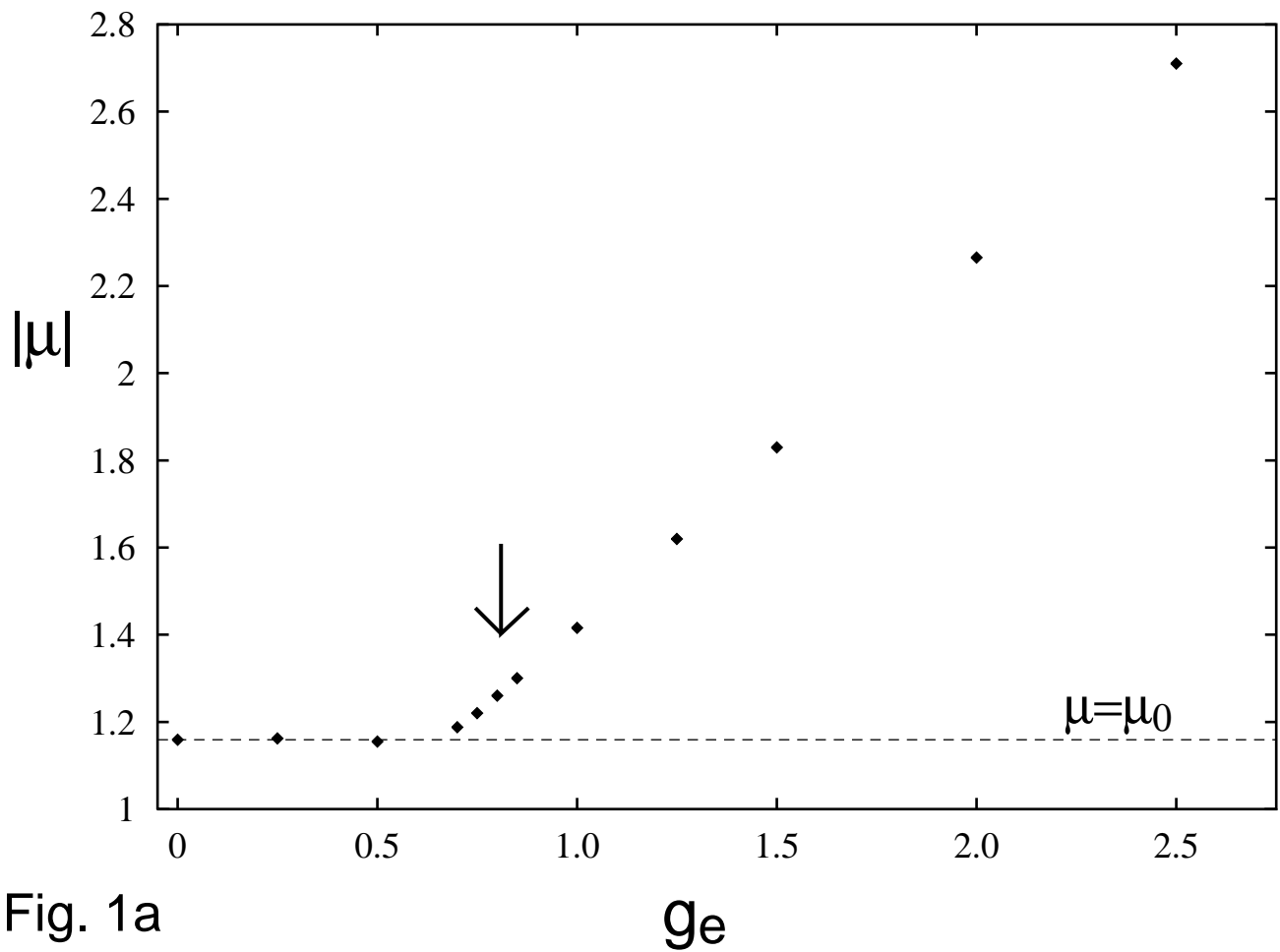
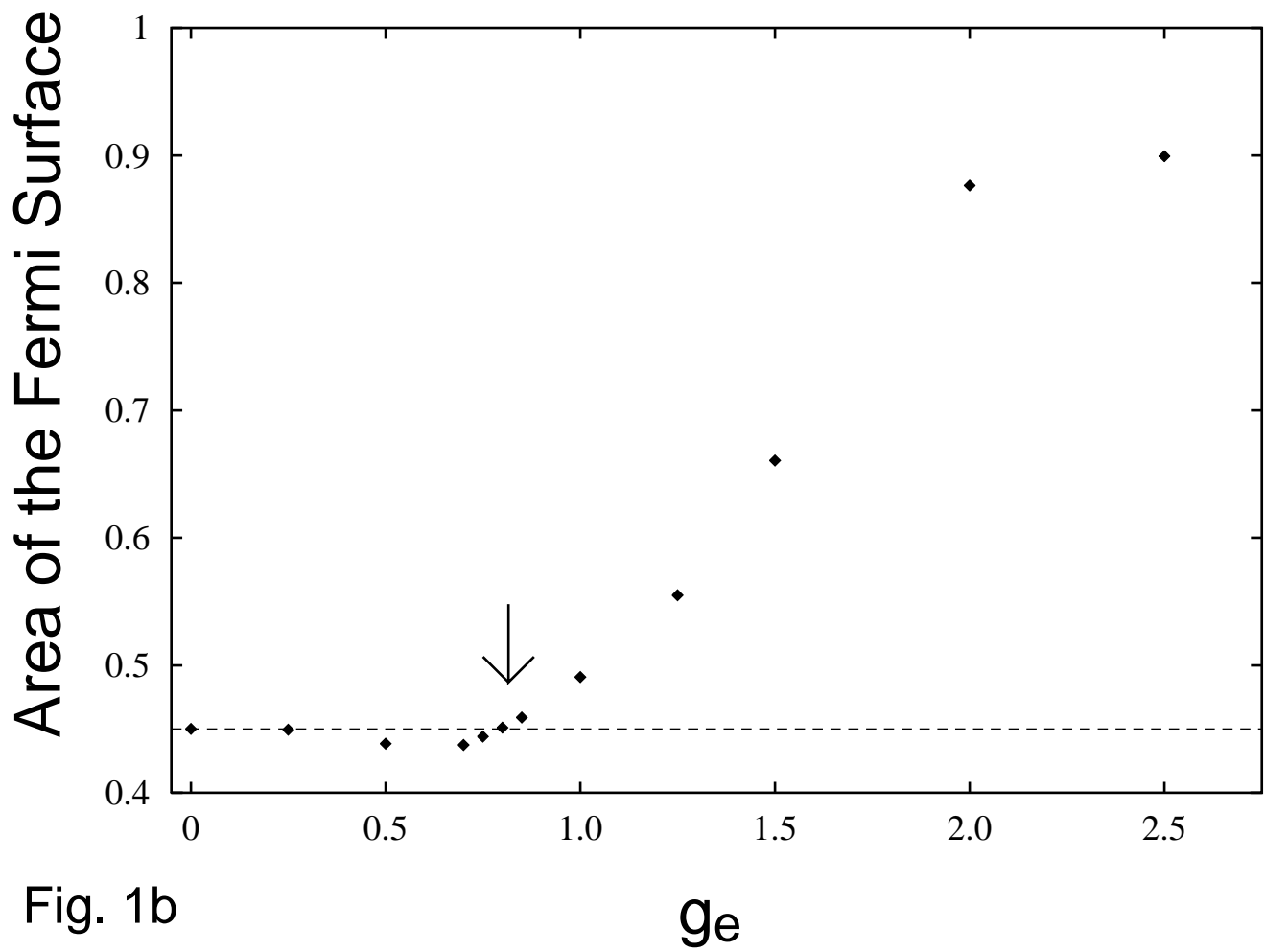
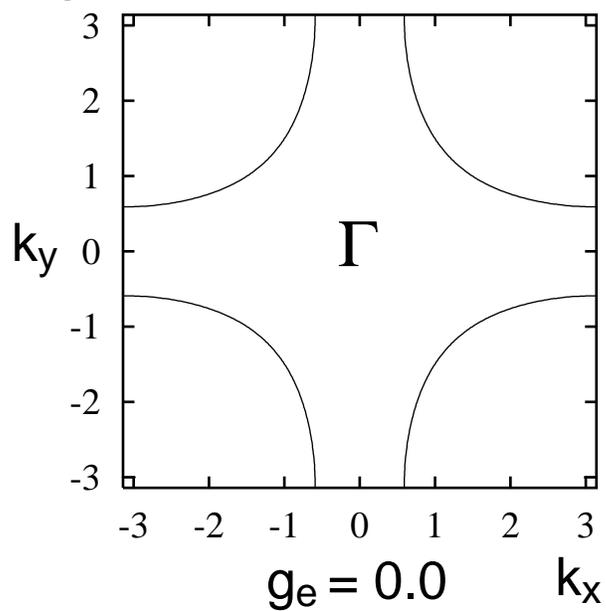
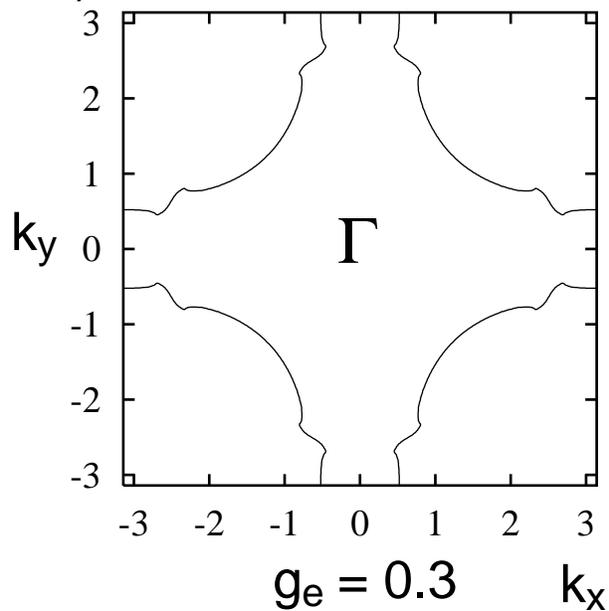
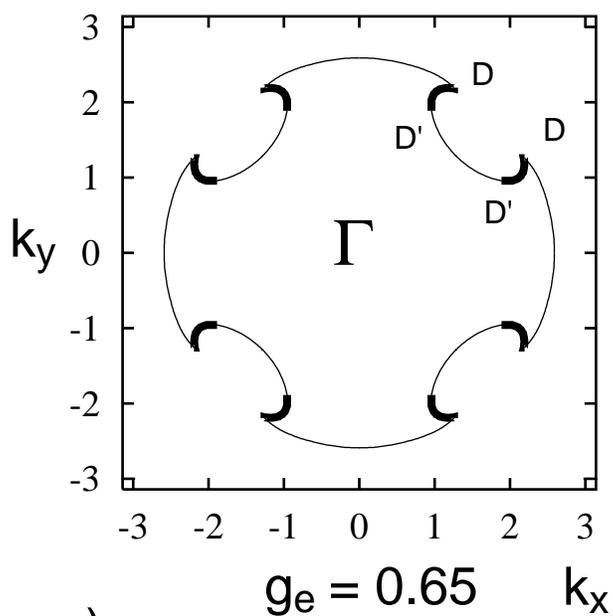
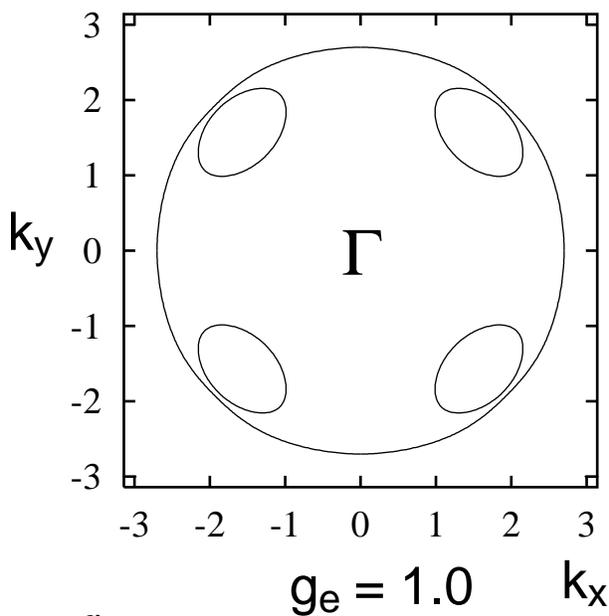
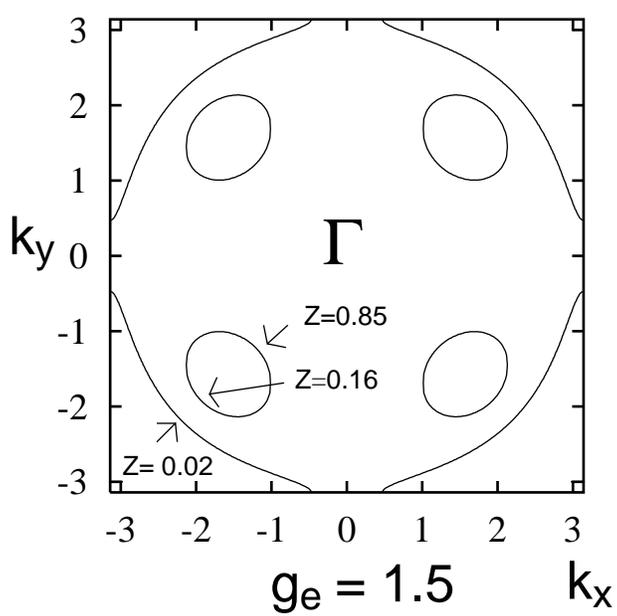
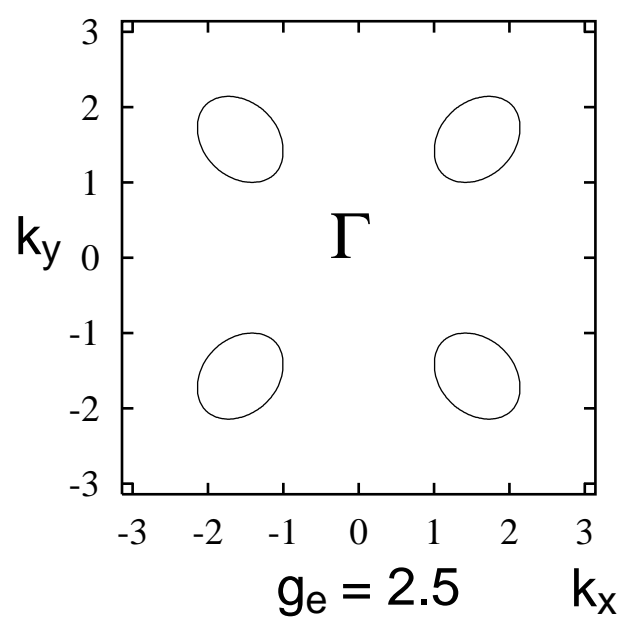
\end{document}